\documentclass[12pt,a4paper]{article}
\usepackage{t1enc}
\usepackage[latin1]{inputenc}
\usepackage[english]{babel}
\usepackage{amssymb,amsmath}
\usepackage{graphicx}
\usepackage{hyperref}
\usepackage{epstopdf}

\begin{document}

\title{The road toward a general relativistic metric inside the Earth and its effect on neutrino travel from CERN to GRAN-SASSO Laboratory.}
\author{Olivier Besida \\  CEA, Irfu, SPP \\  Centre de Saclay, F-91191 Gif-sur-Yvette, France
\\olivier.besida@cea.fr}
\maketitle

\begin{abstract} 
In a first attempt to describe the effect on neutrino travel inside the Earth caused by general relativity in the case of a dense Earth, we have neglected the Earth's rotation, the Earth's ellipticity and also the surface terrain variation, nevertheless we have focused our attention on the density description of the Earth interior provided by geophysic's models such as PREM. Assuming a non rotating Earth, the general relativistic effect on neutrino travelling from CERN to GRAN-SASSO happened to produce a delay of $\delta t=4.1863 \, picosecond$.
\end{abstract}
 
\section{Introduction}
After the paper by the OPERA collaboration (cf. \cite{OPERA}) claiming an advance of the CNGS neutrino beam from CERN to Gran-Sasso  of $(57.8 \pm 7.8 (stat.)+8.3-5.9 (sys.)) ns$, strong excitement occurred around all possible General Relativistic corrections that could account for such a strong effect, leaving a big anomaly on neutrino speed faster than the speed of light of $(v-c)/c = (2.37 \pm 0.32 (stat.) (sys.)) \,10^{-5}$. Two main corrections are to be applied about this measurement:
Firstly the Sagnac Effect(\cite{Ashby}) (not detailed here), which expresses the impact of the earth's rotation around its axis on time synchronization with atomic clocks using GPS in common view mode, and secondly the General Relativistic effect on the neutrino travel due to the presence of the mass of the planet Earth. It is this second effect, that is deeply detailed here, with the implication of a non homogeneous distribution mass density inside the earth. While a simple non-rotating model with spherical symmetry with geophysics density data input, a first calculation is achieved, providing a  $\delta t=4.1888 \, picosecond$ delay effect. Some second order effects are further taken into account such as the ellipticity of the earth, the effect of mountains and of geoid fluctuations, meanwhile the same delay effect of $ \delta t=4.1863 \,picosecond$ is computed. Of course such an effect is far from being measured due to the timing resolution of the most uptodate technics of nowadays neutrino detection and also neutrino beam production with accelerators timing technics. 
Previously Kiritsis and Nitti (cf \cite {Kiritsis}) have calculated the General Relativistic effect of neutrino travel between CERN and Gran-Sasso, but they assumed a metric of a black hole or far from a mass supposed to be ponctual which is not the case here where neutrinos are going inside the earth. Of course if this approximation is slighly wrong at the surface of the earth and on the crust, it is deeply wrong if the neutrino beam is crossing the earth from one side to the other; with how modelization of density distribution inside earth, a delay effect of $86.65 picosecond$ is to be expected for a full crossing along the earth diameter.
 The earth is assumed not to be rotating, so that a Schwartschild metric is sufficient, while in theory a detailed calculation in the frame of a Kerr metric would be needed, nevertheless, this effect is still extremly tiny. This effect cannot account for any advance of the neutrino. Recently the ICARUS experiment also carrying measurement on the CNGS neutrino beam did not notice any discrepancy between the speed of neutrino and the speed of light (cf \cite{ICARUS}).  

\section{Toward a relativistic metric inside the earth}
As we do not take into account the rotation of the earth and a spherically symmetric density model of the earth interior, we can use the Schwartzchild metric to take into account general relativity as follows:\\
\begin{align}
&\ g =(1- \dfrac{2GM}{r c^2})dt^2 -  (1-\dfrac{2GM}{r c^2})^{-1}dr^2 -r^2(d\theta ^ 2+sin^2 \theta d\phi^2)  
\end{align}
The Schwartzchild metric can be easily rewriting using $ \Phi $ potential :
\begin{align}
&\ g =(1- \Phi) dt^2 -(1-\Phi)^{-1} dr^2  -r^2(d\theta ^ 2+sin^2 \theta d\phi^2)  
\end{align}

$ \Phi $ is defined by :
\begin{align}
&\ \Phi = \dfrac{2GM}{r c^2}=-\frac{2V}{c^2} 
\end{align}
V is easily identified as the classical Newtonian gravitational potential:\\
\begin{align}
&\ V=V_{Newton}=-\dfrac{GM}{r} 
\end{align}
At this point, a remark must be made. If Tolman-Oppenheimer-Volkoff equation (TOV) is perfectly suited to describe the interior of a spherically symmetric isotropic interior of a neutron star in isostatic equilibrium between gravity and pressure with strong general relativistic effects with the input of Equation of State (EOS) of the nuclear matter, in the case of the earth the situation is slighlty different. The detailed atomic composition of the inner earth is a big unknown field and the knowledge of its interior (density and pressure) is mainly coming from the long standing experience of geophysicist modelling the seismic waves propagation through the interior of the earth. Then the TOV equation applied to the earth is directly short cut in this work and the density distribution from geophysics is directly plugged into the Schwartzchild metric as an input.

\section{Computing the gravitational potential with Gauss theorem and Poisson equation}

Because of spherical symmetry hypothesis, the density inside the Earth, varies only with r.
While differential formulation of Gauss's law for gravity is given by:
\begin{align}
&\ 4 \pi G \rho = -\vartriangle V
\end{align}
The spherical symmetry implies for the expansion of the Laplacian:
\begin{align}
&\ 4 \pi G \rho = -\frac{1}{r^2} \frac{\partial}{\partial r}(r^2 \frac{\partial V}{\partial r})
\end{align}
The integral form is given by the Gauss theorem :
\begin{align}
&\ V(r)-V_{0} = -4 \pi G \int \limits_{0}^{r} \frac{dr}{r^2} \int \limits_{0}^{r}\rho (r)  r^2 dr
\end{align}
$ M(r) $ is defined as the Earth's mass integrated between the center of the earth and radius $r$.
\begin{align}
&\ M(r) = \int \limits_{0}^{r}  \rho (r) 4 \pi r^2 dr
\end{align}
Gravity acceleration is directly obtained through:
\begin{align}
g(r) &= - \frac{G}{r^2}\int \limits_{0}^{r}  \rho (r) 4 \pi r^2 dr  \nonumber \\
 &=-\frac{GM(r)}{r^2}
\end{align}
The classical Newtonian gravity potential is obtained without difficulty as follows.  
\begin{align}
V(r) &= -\int \limits_{0}^{r} \frac{GM(r)dr}{r^2} +V_0 \nonumber \\
     &= \int \limits_{0}^{r}g(r)dr +V_0
\end{align}

\begin{align}
 \Phi(r) &= -\frac{2V(r)}{c^{2}} \nonumber \\ 
 &= \int \limits_{0}^{r} \frac{2GM(r)dr}{r^2 c^2}+ \Phi _0 \nonumber \\ 
&= -\int \limits_{0}^{r}\frac{2g(r)dr}{c^2} + \Phi _0 \\
&\ \Phi _0 =-\frac{2V_0}{c^2}
\end{align}
The second constant term $V_0$ and $\Phi_0$ in the integral accounts for the prescription of a flat Schwartzchild metric at an infinite distance r from the Earth center, as the newtonian gravitational potential is plugged into Schwartzchild metric through $\Phi$, in such a way that $\Phi(+\infty)=0 $.

\section{The radial density of earth interior obtained by geophysics methods with seismic waves propagation}
\begin{figure}[htp] 
\centering
\includegraphics[width=14cm,height=15cm]{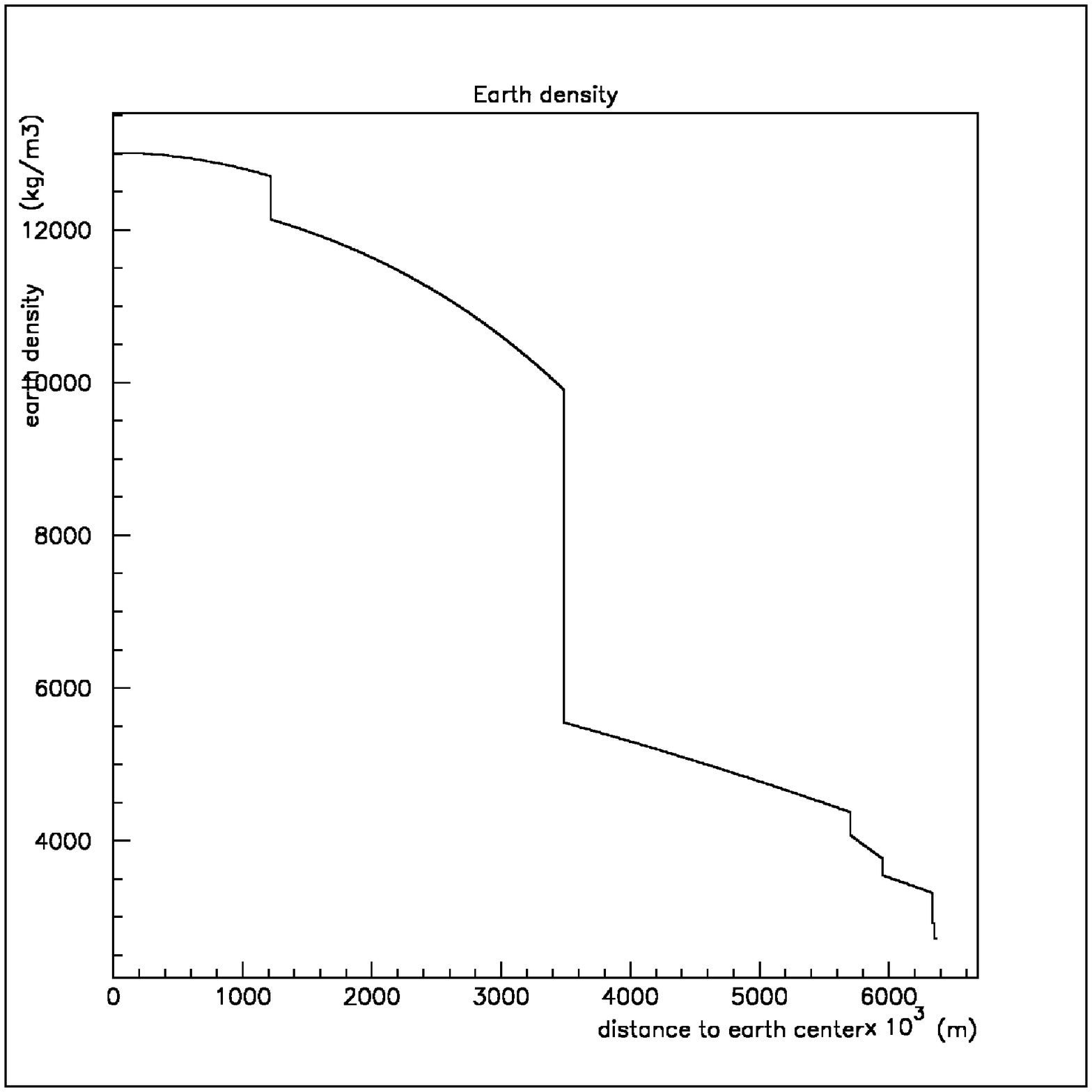}
\caption{Density of the Earth interior (PREMC)} 
\label{fig:densityw} 
\end{figure}

\begin{figure}[htp] 
\centering
\includegraphics[width=14cm,height=15cm]{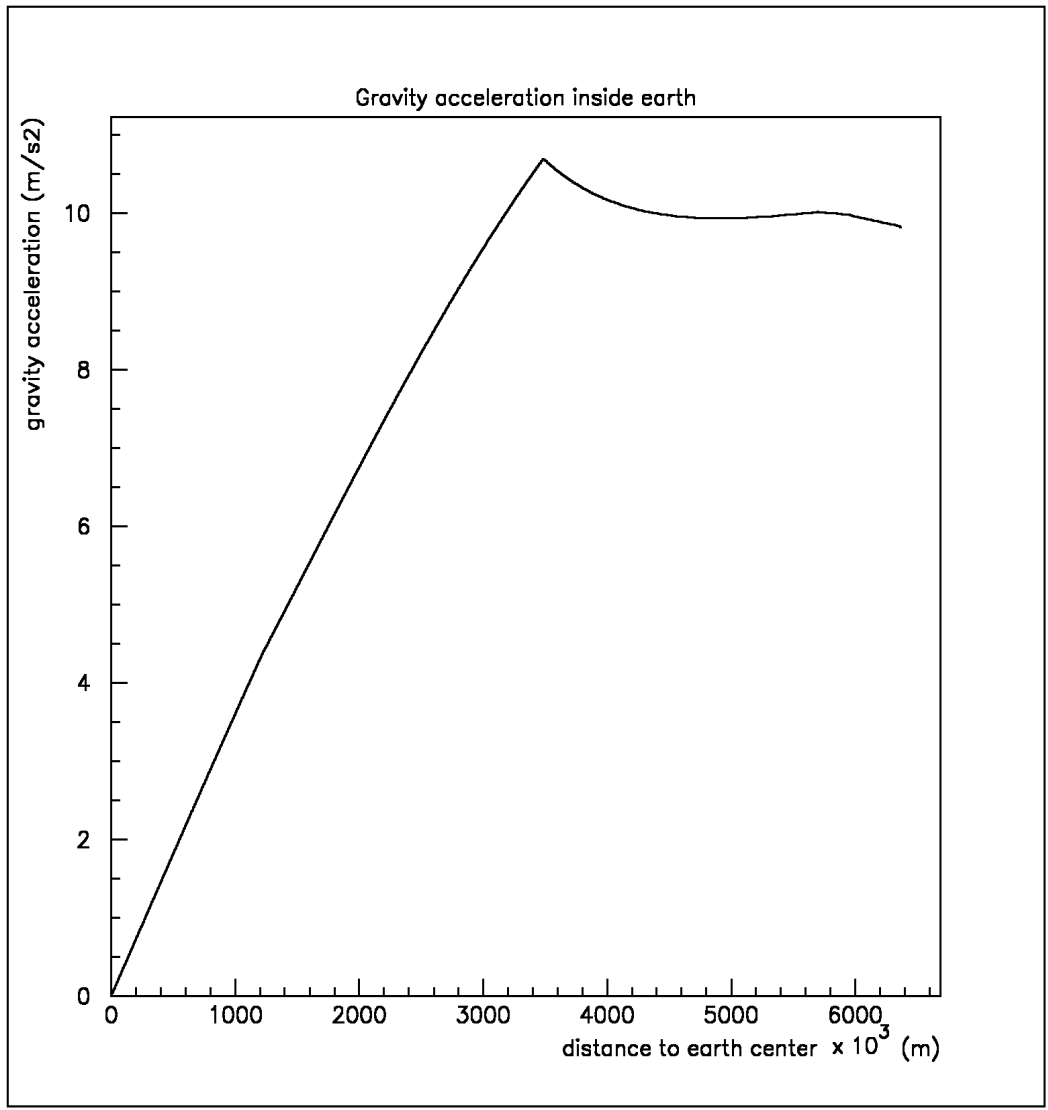}
\caption{Gravity acceleration inside the Earth} 
\label{fig:gravityw} 
\end{figure}

\begin{figure}[htp] 
\centering
\includegraphics[width=14cm,height=15cm]{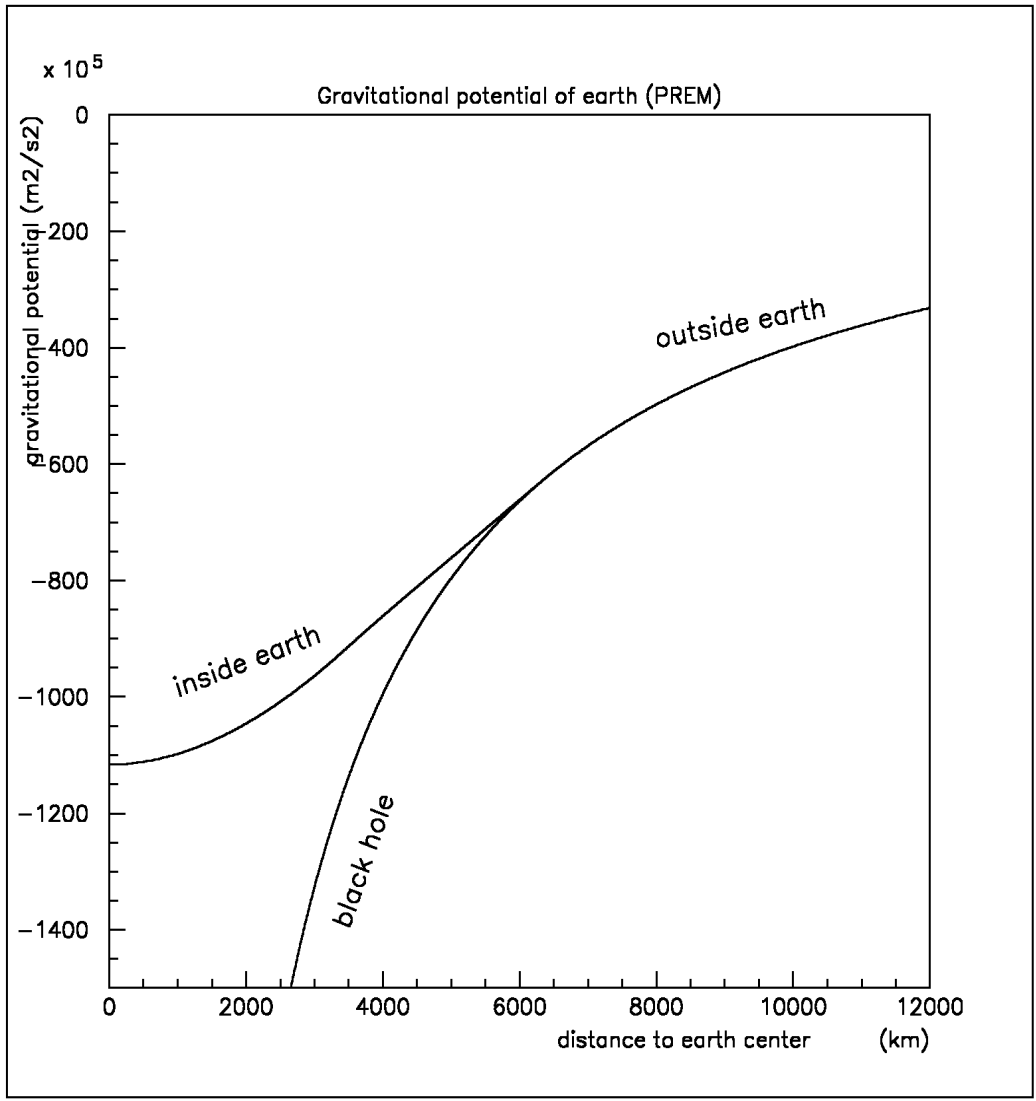}
\caption{Gravitational potential inside the Earth} 
\label{fig:potpremw} 
\end{figure}

\begin{figure}[htp] 
\centering
\includegraphics[width=14cm,height=15cm]{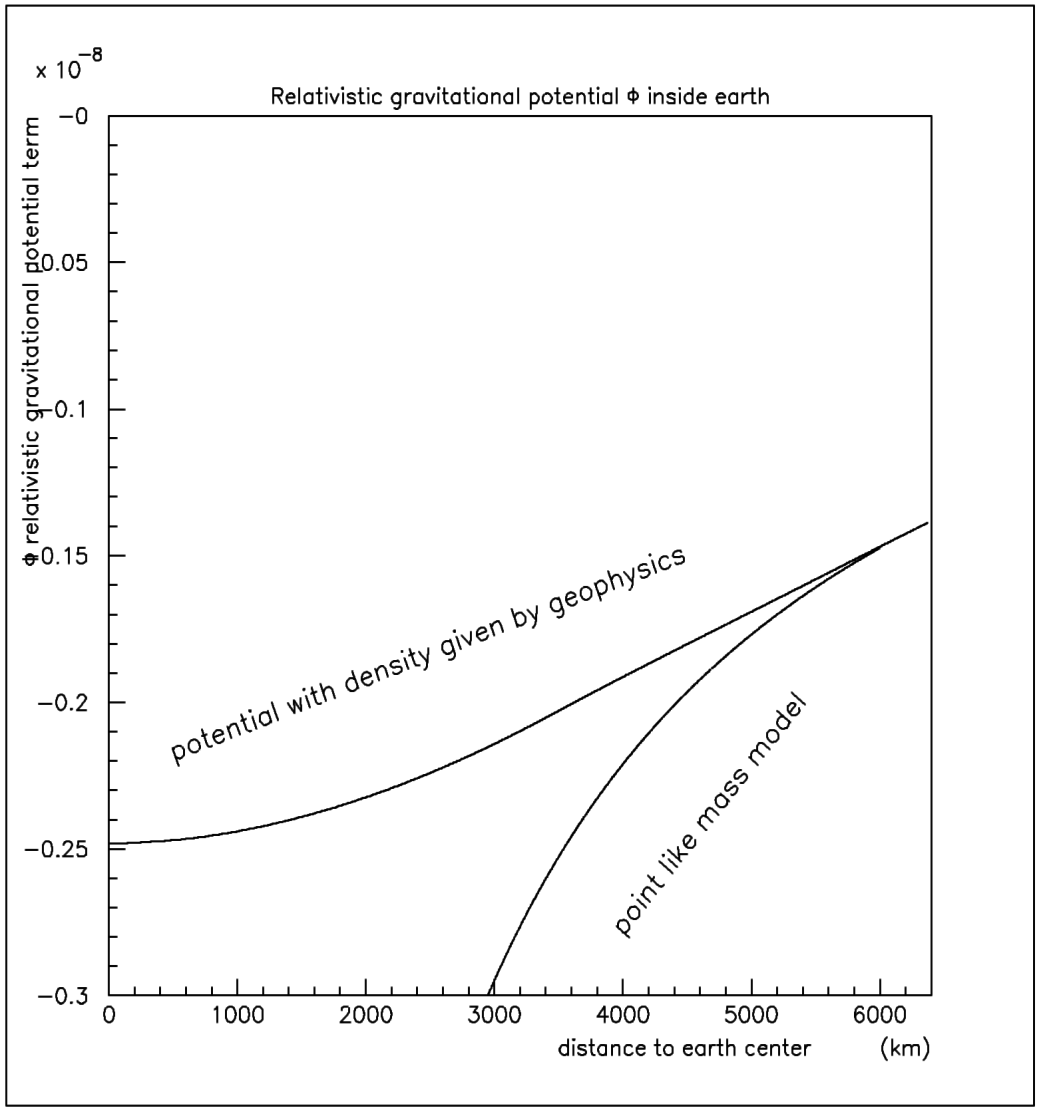}
\caption{General relativistic term of gravitational potential $-\Phi(r) $ inside the Earth compared to a pointlike gravitational potential} 
\label{fig:fipotprem} 
\end{figure}

Since many years, geophysicists have been monitoring earthquakes and seismic waves propagation through the earth interior. Two main kinds of seismic waves crossing the Earth are observed, primary waves (faster) and secondary waves (slower). Beside the difference of speed of these two categories of waves, P-waves are compressional waves, while S-waves are shear waves which do not propagate in liquids. The propagation of these waves inside the Earth and their sophisticated diffraction processes at the interfaces between solid and liquid zone of the inner earth, inform seismologists about the nature of the phase of the matter and its density versus depth inside the Earth. These two speeds can be calculated as follows:
\begin{align}
&\ v_{primary}=\sqrt{\frac{K+(\frac{4}{3})\mu}{\rho}}
\end{align}
\begin{align}
&\ v_{secondary}=\sqrt{\frac{\mu}{\rho}}
\end{align}
Bulk modulus =$ K $\\
Shear modulus $ \mu $\\
Density =$ \rho $ \\
A seismic parameter $\Phi_{seismic}(r)$ is obtained cancelling the shear modulus out of the primary and the secondary waves.
\begin{align}
&\ \Phi_{seismic}(r)= v_{primary}^{2}-\frac{4}{3} v_{secondary}^{2}
\end{align}
Moreover, assuming  hydrostatic equilibrium and adiabatic exchanges, Adams and Williamson (1923) showed a relationship between this geophysic parameter $ \Phi_{seismic} $, the density and the gravity acceleration at a distance r from the center of the Earth:
\begin{align}
&\ \frac{d \rho}{dr} =- \dfrac{\rho(r)g(r)}{\Phi_{seismic}(r)}  
\end{align}
The numerical integration of the Adams and Willimason equation with $\Phi_{seismic}(r)$ gives the density (see Figure ~\ref{fig:densityw} ).
Then we computed the gravity acceleration versus r (see Figure ~\ref{fig:gravityw} ) and afterward the Newton gravitational potential inside earth and so the $\Phi$ potential term of the Schwartzchild metric inside the Earth (see Figure ~\ref{fig:potpremw}).
The $\Phi_{seismic}(r)$ input data are obtained with a modelization of the propagation of primary and secondary seismic waves for any depthes inside the Earth in order to reproduce the observed seismic waves patterns measured by seismographs scattered all over the earth surface.
Each seismograph is recording a seismic wave train characteristic of the angle between its position and the epicenter of each earthquake under study, and also of the depth of the epicenter below the surface. Nowadays, sophisticated models exist (PREM, PEM,IASP91) after the first attempts by Bullen (cf. \cite  {Bullen}) and others (1937). 
While the most important part of the surface of the Earth is covered by ocean, a continental Model PREMC does exist and we have used this model with a density of $2.72 tons/m^{3}$ for the upper continental crust instead of $1. tons/m^{3}$ for the ocean.
We focused our attention on the Preliminary Earth Reference Model (PREM) by Dziewonski \& Anderson (1981) (cf. \cite{DZIEWONSKI}) which is highly sufficient for our purpose of integrating the density along the radius of the earth in order to obtain firstly the acceleration of the gravity, secondly the newtonian gravitational potential inside the earth and thirdly a spherically symmetric Schwartzchild-like metric inside the earth ( cf. \cite{SAXOV}, \cite{Ignatiev} \& \cite{PEC}). (see Figures ~\ref{fig:gravityw}, ~\ref{fig:potpremw} ~\ref{fig:fipotprem})
\begin{align}
&\ r=R/6371. km
\end{align}

PREM shows various zones inside earth: \\

Inner Core (solid):
\begin{align}
 0. km <& R < 1217.1 km \\ 
 0. <& r < .19103 \\
 \rho(r) =& 13.0122-8.45292r^{2} \\
\Phi(r) =& 2.4845\,10^{-9}-1.6396\,10^{-9}r^{2} \nonumber \\
& +3.1953\,10^{-10}r^{4}
\end{align}
Outer Core (liquid):
\begin{align}
 1217.1 km <& R < 3458.7 km \\ 
 .19103 <& r < .54712 \\
 \rho(r) =& 12.5842 -1.69929r-1.94128r^{2} -7.11215r^{3} \\
\Phi(r) =& 2.4950\,10^{-9}-1.3621\,10^{-10}r-1.0558\,10^{-9}r^{2} \nonumber \\
& -1.0388\,10^{-9}r^{3}+1.3768\,10^{-9}r^{4}-4.3678\,10^{-10}r^{5}
\end{align}
Lower Mantle:
\begin{align}
3458.7 km <& R < 5701. km \\
.54288 <& r < .89483 \\
\rho(r) =& 6.8143 -1.66273r -1.18531r^{2}\\
\Phi(r) =& 3.3847\,10^{-9}-4.2421\,10^{-9}r+5.1207\,10^{-9}r^{2} \nonumber \\
& -4.0832\,10^{-9}r^{3}+1.2106\,10^{-9}r^{4}
\end{align}
Transition Zone:
\begin{align}
5701. km <& R < 5951. km \\ 
.89483 <& r < .93407 \\
\rho(r) =& 11.1198 -7.87054r\\
\Phi(r) =& 2.1071\,10^{-9}+6.0380\,10^{-10}r-1.63960\,10^{-9}r^{2} \nonumber \\
& +3.19532\,10^{-10}r^{4}  
\end{align}
Low Velocity Zone:
\begin{align}
5951. km <& R < 6336. km \\ 
.93407 <& r < .99450 \\
\rho(r) =& 7.15855 -3.85999r \\
\Phi(r) =& 2.7521\,10^{-9}-8.0668\,10^{-10}r-1.4230\,10^{-9}r^{2} \nonumber \\
& +1.2167\,10^{-9}r^{3}-3.4782\,10^{-10}r^{4}
\end{align}
Lower Crust:
\begin{align}
6336. km <& R < 6351. km \\ 
.99450 <& r < .99686 \\
\rho(r) =& 2.92 \\
\Phi(r) =& 3.5782\,10^{-9}-2.3585\,10^{-9}r-1.4790\,10^{-10}r^{2} \nonumber \\
& +3.1953\,10^{-10}r^{4}
\end{align}
Upper Crust:
\begin{align}
6351. km <& R < 6371. km \\ 
.99686 <& r < 1. \\
\rho(r) =& 2.72 \\
\Phi(r) =& 3.6310\,10^{-9}-2.4644\,10^{-9}r-9.4700\,10^{-11}r^{2} \nonumber \\
& -3.1953\,10^{-10}r^{4}
\end{align}

\section{Integrating the travel time of the neutrino along a geodesic inside earth}
\subsection{Shapiro delay inside earth}
In the sixties, I. Shapiro proposed a test of General Relativity using the delay of radar bouncing over Mercury and Venus checking the relativistic effect due to the mass of the sun as the radar beam towar these planets was coming closed to the sun, this is known as time delay of radar echoes (cf \cite {Shapiro1}, \cite {Shapiro2}, \cite {Straumann}).
The Lagrangian of the trajectory of the neutrino inside earth:
\begin{align}
&\ \mathcal{L}=\frac{1}{2} g_{\mu \nu} \dot{x} ^{\mu} \dot{x} ^{\nu}
\end{align}
Inserting the metric into the Lagrangian:
\begin{align}
&\ 2\mathcal{L}=(1- \Phi) \dot{t} ^2 -(1- \Phi)^{-1}\dot{r} ^2 - r ^2(\dot{\theta} ^2+sin^2 \theta \dot{\phi}^2))
\end{align}

Along the orbit, the Lagrangian $\mathcal{L}$ is constant. A simple change of coordinates can be applied with the condition $\theta  = \frac{\pi}{2} $ which implies : $ \dot{\theta}=0$

\begin{align}
&\  \Rightarrow 2\mathcal{L}=(1- \Phi) \dot{t} ^2 -(1- \Phi)^{-1}\dot{r} ^2 - r ^2 \dot{\phi}^2
\end{align}
Applying Noether's Theorem to the invariance of this Lagrangian (and of the $ \Phi $ potential) respect to the variables $\phi $ and $ t$ , two conserved quantities are extracted $ L $ and $ E $ as follows :
\begin{align}
&\ -\frac{\partial \mathcal{L}}{\partial \dot{\phi}} = r ^2 \dot{\phi} =L =\textit{cte}
\end{align}
\begin{align}
&\ -\frac{\partial \mathcal{L}}{\partial \dot{t}} =  \dot{t} (1- \Phi) =E =\textit{cte}
\end{align}
After some simple change in the equation, we get:
\begin{align}
&\ \dot{\phi} = \frac{L}{r ^2}
\end{align}
\begin{align}
&\ \dot{t} = \frac{E}{(1- \Phi)}
\end{align}
Neglecting the mass of the neutrino (few milli-eV) respect to its energy (17 GeV), the Lagrangian is set equal to zero as in the photon case:
\begin{align}
&\ \Rightarrow 2\mathcal{L}=(1- \Phi)^{-1} (E^2 -\dot{r} ^2) - \frac{L^2}{r ^2}=0
\end{align}
Simplified as:
\begin{align}
&\ \dot{r} ^2 = E^2 -(1- \Phi)\frac{L^2}{r ^2}
\end{align}
The derivation of $r$ respect to the time is obtained here:
\begin{align}
&\ \dot{r}=(\frac{dr}{dt})\dot{t}=(\frac{dr}{dt})E(1- \Phi)^{-1}
\end{align}
Thus:
\begin{align}
&\ (\frac{dr}{dt})^2 E^2 (1- \Phi)^{-2} = E^2 -(1- \Phi) \frac{L^2}{r ^2}
\end{align}
Thanks to the spherical symmetry of this central potential, the neutrino crosses the closest point to the center of the earth at a distance $r=r_{0}$; this distance is a minimum, then $\frac{dr}{dt}=0$. At this minimum distance, the potential is also an extremum, this value is assigned to $\Phi=\Phi_{0}$. 
\begin{align}
&\ \Rightarrow  E^2-(1- \Phi_{0}) \frac{L^2}{r_{0} ^2} = 0
\end{align}
\begin{align}
&\ \Leftrightarrow (\frac{L}{E})^2=\frac{r_{0}}{1- \Phi_{0}}
\end{align}
L and E are replaced by  $r_{0}$ and $\Phi_{0}$ :
\begin{align}
&\ \Rightarrow (\frac{dr}{dt})^2 (1- \Phi)^{-2}= 1-\dfrac{1- \Phi}{1- \Phi_{0}}(\dfrac{r_{0}}{r})^2
\end{align}
\begin{align}
&\  dt=\dfrac{dr}{ (1- \Phi) \sqrt{ 1-\dfrac{1- \Phi}{1- \Phi_{0}}(\dfrac{r_{0}}{r})^2} } 
\end{align}
The duration of the travel from the surface (point A) to the minimum distance $r_{0}$ is calculated through this integral:
\begin{align}
&\  t_{A}-t_{0}= \int\limits_{r_{0}}^{r_{A}} \dfrac{dr}{ (1- \Phi) \sqrt{ 1-\dfrac{1- \Phi}{1- \Phi_{0}}(\dfrac{r_{0}}{r})^2}} 
\end{align}
The total duration of the travel of the neutrino from point A to point B is computed adding the two integrals:
\begin{align}
&\  t_{B}-t_{A}= \lvert t_{B}-t_{0}\rvert +\lvert t_{A}-t_{0}\rvert \\
&\  t_{B}-t_{A}= \lvert \int\limits_{r_{0}}^{r_{B}} \dfrac{dr}{ (1- \Phi) \sqrt{ 1-\dfrac{1- \Phi}{1- \Phi_{0}}(\dfrac{r_{0}}{r})^2}} \rvert + \lvert \int\limits_{r_{0}}^{r_{A}} \dfrac{dr}{ (1- \Phi) \sqrt{ 1-\dfrac{1- \Phi}{1- \Phi_{0}}(\dfrac{r_{0}}{r})^2}} \rvert
\end{align}

\subsection{Numerical application with PREM based metric}
\begin{figure}[htp] 
\centering
\includegraphics[width=14cm,height=15cm]{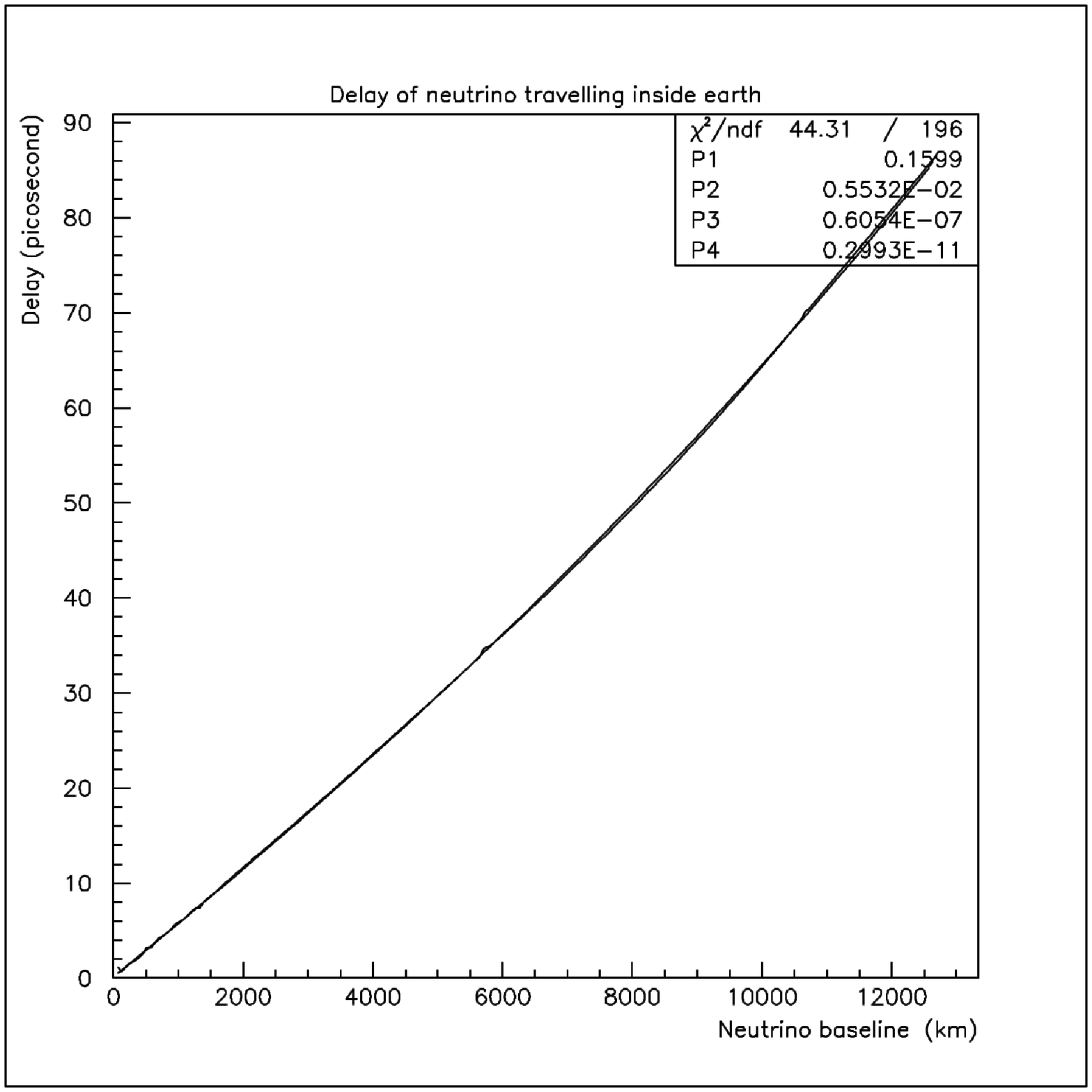}
\caption{Neutrino delay of arrival time versus neutrino baseline length} 
\label{fig:delayvsbaselinew} 
\end{figure}

\begin{figure}[htp] 
\centering
\includegraphics[width=14cm,height=15cm]{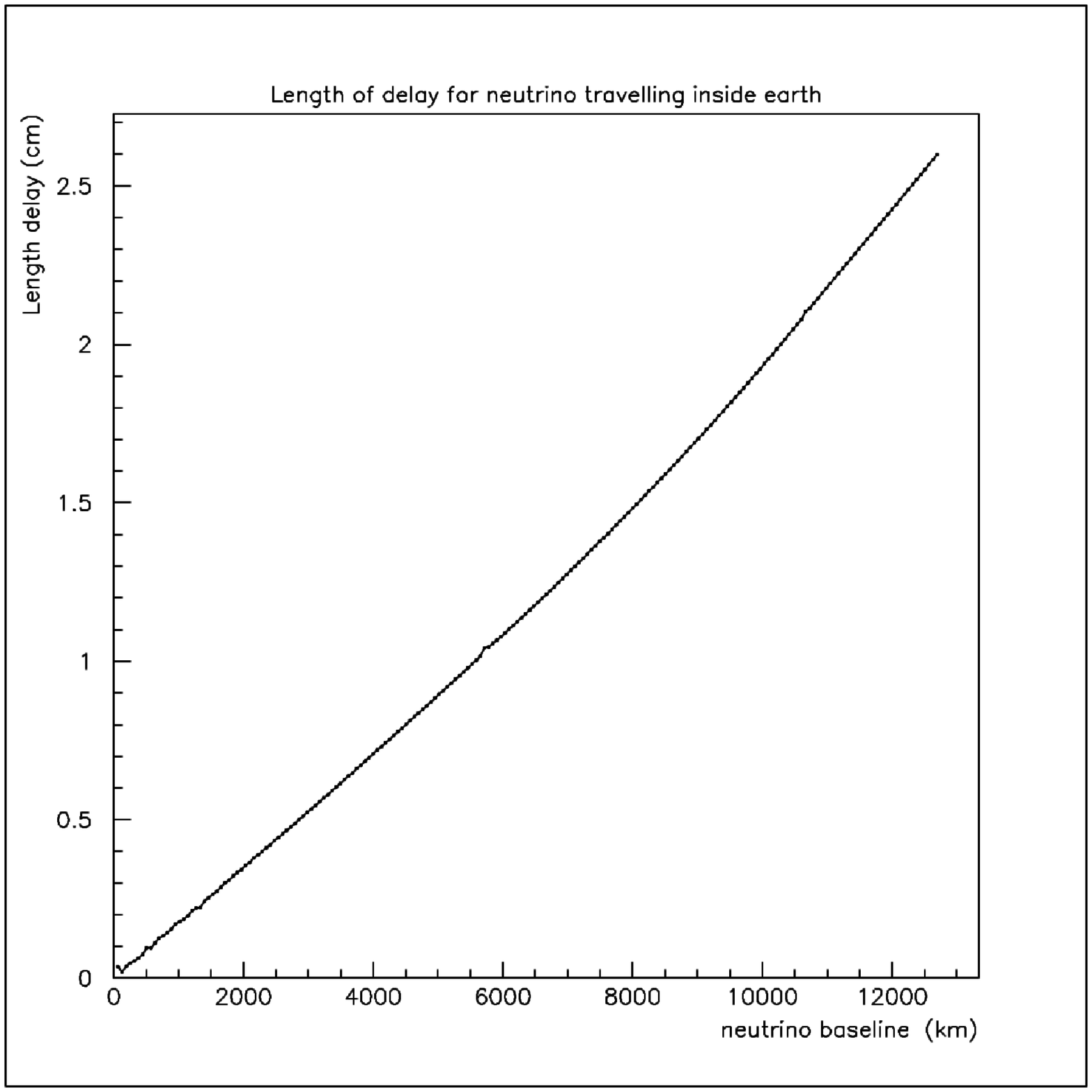}
\caption{Neutrino length delay of arrival time versus neutrino baseline} 
\label{fig:lengthdelayq} 
\end{figure}

The computation of the general relativistic effect for a neutrino baseline of 730 km leads to a delay of $ 4.1888 \, picosecond $, which is equivalent to a delay length of $ 1.255 \, mm $ . In other words the effect decreases the apparent speed of light of a relative amount $\frac{\delta c}{c} = 1.7199 \,  10^{-9} $.
This computation is obtained with a Monte-Carlo integration method applied to equation (63) with $10^{9} $ iterations.
We have computed this effect for 200 distances of neutrino baseline crossing the earth spanning between 63 km to 12700 km each one based on the same Monte-Carlo method (see figures ~\ref{fig:delayvsbaselinew} , ~\ref{fig:lengthdelayq}).
The computed result, in the hypothetical case that a neutrino beam crosses the whole earth diameter (12700km), is a global delay effect of $ 86.65 \, picosecond$, which is equivalent to a length $= 2.599 \, cm$. This delay length is a one "inch effect". Also the apparent speed of light (neutrino speed) is decreased of a relative amount  $\frac{\delta c}{c} =2.046 \, 10^{-9} $.
In any case, no advance compared to the speed of light is seen here, only a delay through this general relativistic effect due to the gravitational potential caused by the mass distribution inside the Earth.
Anyhow, this calculation is done with the following assumptions, no rotation of the earth, the earth is supposed to be perfectly spherical and the computation is done from one point of the surface to another point of the surface with neither mountain, nor any relief, with neither an ellipsoid shape of the earth, nor a geoid frame of the earth (i.e.  without any angular anisotropic distribution of masses inside the earth).
A fit of the computed delay $\delta t [picosecond]$ versus the neutrino baseline length $b  [km]$ is expressed as follows:
\begin{align}
&\ \delta t  [ps]=.1599+.55318 \, 10^{-2}b+.60543 \, 10^{-7}b^{2}+.29928  \, 10^{-11}b^{3} 
\end{align}

\section{Further improvements beyond this simple spherical model}

\subsection{Effect of the ellipticity of the earth within the WGS84 frame}
The ellipticity of the earth is fully described with the World Geodetic System 84 (WGS84) (see: \cite {NIMA}), this geodetic system describes an ellipsoid as a good approximation of the Earth surface. The origin of the ellipsoid is the center of the earth. This ellipsoid is characterised by two primary parameters $a$ and $e$: \\ 
Semi-major axis = a = 6378137. m \\
Eccentricity = $e = \frac{\sqrt{a^2-b^2}}{a} =0.081819190842622$ \\
There are also some secondary parameters:\\
Semi-minor axis = b = 6356752.3142 m \\
Flattening = $f = \frac{a-b}{a} = 0.00335281066474$\\
Any position on this ellipsoid can be referenced with the longitude, geocentric latitude (angle refered to earth center) or geodetic latitude (angle refered to the normal of the ellipsoid) and radius, or in cartesian (x,y,z) coordinates.\\
Geodetic latitude = $\phi $ \\
Geocentric latitude = $\phi^\prime$ \\
Longitude = $\lambda $ 
The relationship between the geodetic and the geocentric latitude are defined as follows:
\begin{align}
&\ tan(\phi^\prime) = ( \frac{b}{a})^{2} tan(\phi)
\end{align}
The relationship between the cartesian position (x,y,z) and the latitude, longitude plus radius cordinates are given by:
\begin{align}
&\ x=\frac{a cos (\phi) cos (\lambda) }{\sqrt{1-e^{2}sin^{2} (\phi) }} \\ 
&\ y=\frac{a cos (\phi) sin (\lambda) }{\sqrt{1-e^{2}sin^{2} (\phi) }} \\
&\ z=\frac{a(1-e^{2}) sin( \phi) }{\sqrt{1-e^{2}sin^{2} (\phi) }}  \\
&\ r_{ellipsoid}(\phi)=a\sqrt{1-\dfrac{e^{2}(1-e^{2})sin^{2} (\phi)}{1-e^{2}sin^{2} (\phi)}}
\end{align}
The value of the gravity acceleration at the surface of the ellipsoid is easily computed with the geodetic latitude :
\begin{align}
&\ \gamma (\phi) =  \gamma_{equ.} \dfrac{(1+ k sin^{2}( \phi))}{\sqrt{(1 - e^{2}sin^{2}( \phi))}}
\end{align}
Normal gravity at the equator = $\gamma_{equ.} = 9.7803253359 \ \ m/s^{2}$
\begin{align}
&\ \gamma (\phi) = (9.7803253359)  \dfrac{(1+0.00193185265241 sin^{2} (\phi))} { \sqrt{(1-0.00669437999014 sin^{2} ( \phi))}} \ \  m/s^{2}
\end{align}

\subsection{Positions of the trajectory of neutrino from CERN to Gran-Sasso in ECEF WGS84 frame}
In the Earth Centered, Earth Fixed (ECEF) coordinate system, the position of CERN reference point( T-40-S-CERN) of the CNGS neutrino beam is defined as:\\
\begin{em}
x $=4394369.327 $  y $=467747.795 $ z $=4584236.112$ \\
Latitude =46 \textdegree 14 $'$ 32.793 $"$ N Longitude =6 \textdegree 4 $'$ 32.987 $"$ E \\
Altitude\, referred \, to \, earth \, ellipsoid \, (WGS84)$ = 429.72 \, m$\\
Distance \, to \, earth \, center$ = 6367455.61\, m $\\
 \\
\end{em}
The Opera reference point (A1-9999) is defined as:\\
\begin{em}
x $ =4582167.465$ y $ =1106521.805 $ z $ =4283602.714 $\\
Latitude = 42 \textdegree 27 $'$13.324 $"$ N Longitude = 13 \textdegree 34 $'$ 34.035 $"$ E \\
Altitude referred  to  earth  ellipsoid \, (WGS84)$ = 1012.22\, m$\\
Distance  to  earth  center$ = 6369450.50 \,m$ \\
 \\
\end{em}
The middle of the neutrino trip between CERN and OPERA reference points is located in the North of Tuscany, in the Parco Nazionale dell'Appenino Tosco Emiliano, close to the village of Casalina, near the towns of Potremoli and Montelugno :\\
\begin{em}
Latitude = 44 \textdegree 24 $'$ 34.289 $"$ N  Longitude = 9 \textdegree 46 $'$ 49.737 $"$ E \\
Depth referred  to  earth  ellipsoid (WGS84) $= 9740.46 \, m $ \\
Distance  to  earth  center$ = 6357969.42 \, m$ \\
Radius  of  ellipsoid   at   the  surface  (WGS84)$=6367709.88 \, m$  \\
 \\
\end{em}
The point of the closest approach of the CNGS neutrino beam to the center of the earth is located  near Barzana di Sotto at :\\
\begin{em}
Latitude =   44 \textdegree 29 $'$ 59.289 $"$ N Longitude = 9 \textdegree 46 $'$ 5.733 $"$  E \\
Depth referred  to  earth  ellipsoid  (WGS84) $= 9730.56 \, m$ \\
Distance  to  earth  center $= 6357945.63 \, m $  \\
Radius  of   ellipsoid   at   the   surface   (WGS84) $= 6367676.20 \, m $  \\
 \\
\end{em}
The coordinates of the position where the CNGS neutrino beam is exiting from the earth toward the sky is located on Campo Imperatore not far from a crossroad toward Santo Stefano di Sessanio, at a distance of 9750m from Opera reference point:\\
\begin{em}
Latitude =   42 \textdegree  24 $'$ 0.278 $"$ N Longitude  = 13 \textdegree  40 $'$ 10.592 $"$  E  \\
Altitude referred  to  earth  ellipsoid  (WGS84) $ = 1585.49 \, m$ \\
Distance  to  earth  center  $= 6370043.69 \, m$ \\
Radius  of   ellipsoid   at   the   surface   (WGS84) $= 6368458.20 \, m$ 
\end{em}
\subsection{Effect of the GEOID (EGM08) or the angular fluctuations of the gravitational potential}
\begin{figure}[htp] 
\centering
\includegraphics[width=14cm,height=15cm]{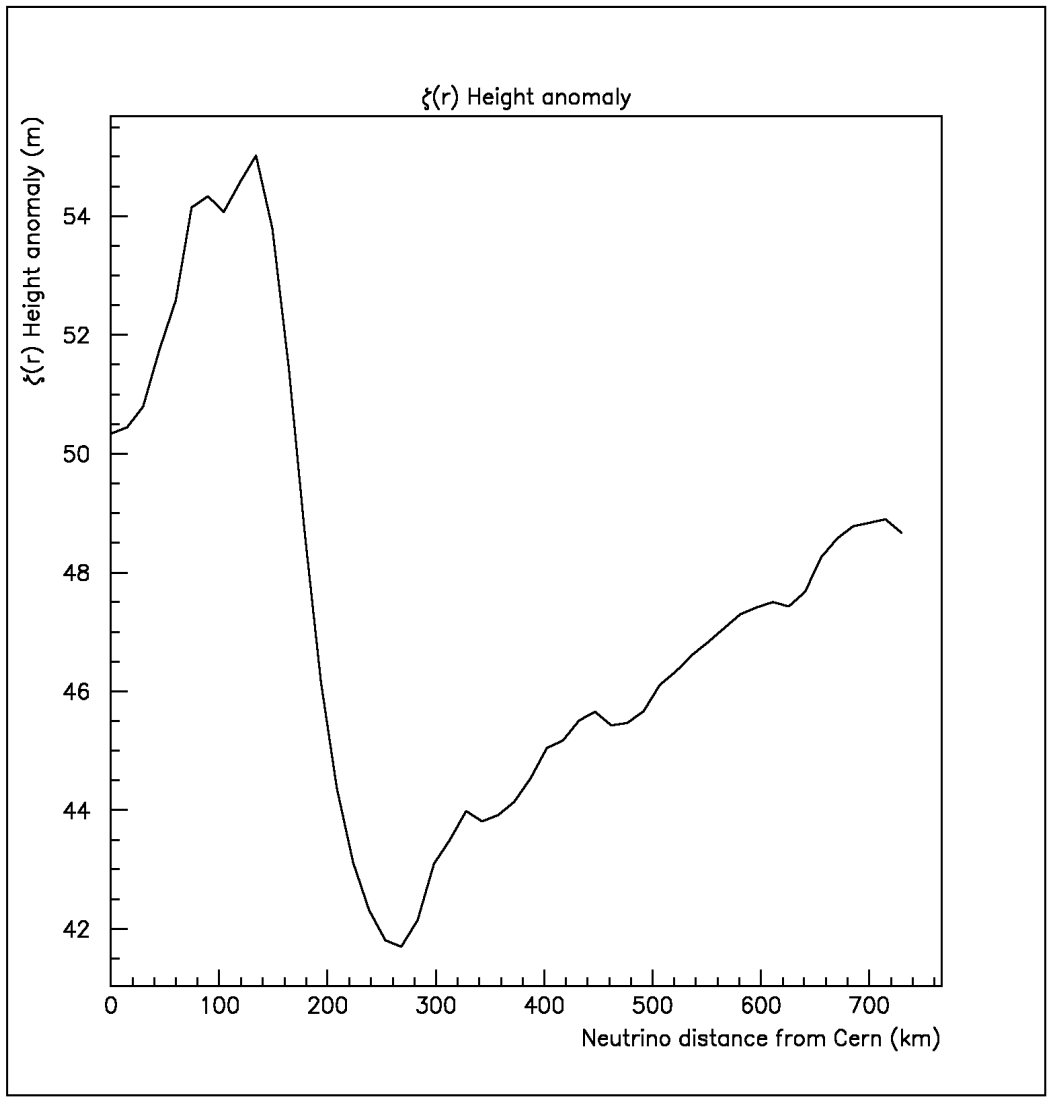}
\caption{Height anomaly versus distance of travel from CERN to Gran-Sasso Laboratory} 
\label{fig:heightanomaly} 
\end{figure}

The geoid undulation of the geoid reflects the fluctuation in height around the surface of the reference ellipsoid (WGS84 here) which defines an equipotential surface of gravity. The mean sea level is a good illustration of this geoid surface. (cf. \cite {NIMA},  \cite{Pavlis})
\begin{align}
&\ V(r,\lambda, \phi^\prime) =-\frac{GM}{r} [1+\sum_{n=2}^{n_{max}}\sum_{m=0}^{n} {(\frac{a}{r})}^n \bar{P}_{nm}(sin(\phi^\prime))(\bar{C}_{nm} cos(m\lambda)+\bar{S}_{nm} sin(m\lambda))]
\end{align}
Normalized associated Legendre function $= \bar{P}_{nm}$ \\
Normalized gravitational coefficients  $=\bar{C}_{nm},\bar{S}_{nm} $ \\
EGM96 nmax=360, resolution $=1\,^{ \circ} \equiv $ 60 nautical miles= 111.12 km\\
EGM08 spherical harmonic degre and order 2159, resolution$ = 6 \prime \equiv$ 10 nautical miles= 18.52 km\\
Height anomaly $= \zeta(r,\lambda, \phi^\prime)$
\begin{align}
&\ \zeta(r,\lambda, \phi^\prime) =\frac{GM}{\gamma(\phi^\prime)r} [\sum_{n=2}^{n_{max}}\sum_{m=0}^{n} {(\frac{a}{r})}^n \bar{P}_{nm}(sin(\phi^\prime))(\bar{C}_{nm} cos(m\lambda)+\bar{S}_{nm} sin(m\lambda))]
\end{align}
We have calculated the value of the gravitational potential at the surface ellipsoid (WGS84) using the value of the height anomaly in EGM08, thanks to the  program $  hsynth \_ WGS84.f $ EGM2008 Harmonic Synthesis, provided by NGA, Geodesy and Geophysics Basic and Applied Research  (\cite{Pavlis}) (see figure ~\ref{fig:heightanomaly}).
\begin{align}
&\ V(r,\lambda, \phi^\prime) = -\frac{GM}{r} - \zeta(r,\lambda, \phi^\prime)\gamma(\phi)
\end{align}
\subsection{Bouguer and free air anomaly corrections of the gravity inside mountains above the ellipsoid}
\begin{figure}[htp] 
\centering
\includegraphics[width=14cm,height=15cm]{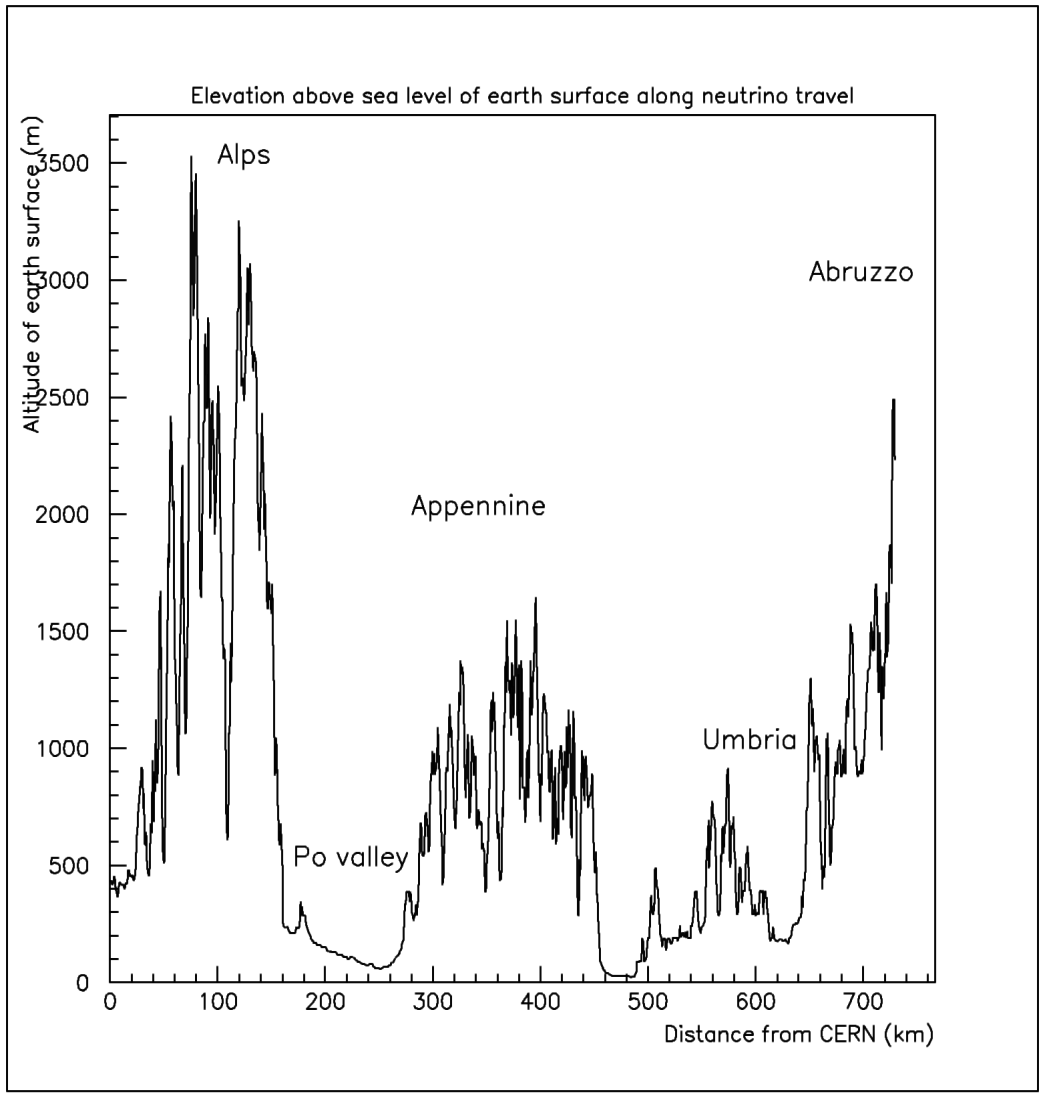}
\caption{Height of the Earth surface versus distance of travel from CERN to Gran-Sasso Laboratory above sea level from GTOPO30 (Credit: U.S. Geological Survey Department of the Interior/USGS)} 
\label{fig:height} 
\end{figure}

There are two additional gravity corrections to take into account: the Free-air anomaly correction and the Bouguer anomaly correction.
The Free-air anomaly is an altitude correction of the gravity.
The Bouguer anomaly corrects the gravity from the presence of mountains that adds some masses close to the point of measurement of the gravity. To take this effect into account, it is assumed that an infinite horizontal plate of thickness h and density $\rho$ is present between the ellipsoid and the top of the mountain where the gravity is supposed to be measured.(cf. \cite {NGIA} \& \cite {NIMA}).
While the measurement is made in a mountain above the ellipsoid but below the surface of the mountain, the Free-air anomaly and Bouguer anomaly are given as follows:
\begin{align}
\Delta g_{F} &= g- \gamma + 4\pi\rho Gd - (\frac{\partial \gamma}{\partial h})(h-d) - (\frac{1}{2})(\frac{\partial^2 \gamma}{\partial h^2})(h-d)^2 +\delta g_{A} \\
4\pi G\rho d &=2.238 \,10^{-6}d
\end{align}
Free air gravity anomaly = $ \Delta g_{F}$ \\
Observed gravity = $ g $\\
theoretical gravity at the level of the ellipsoid = $ \gamma$\\
height of the surface of the earth above the ellipsoid = $ h $ \\
depth of the point under study below the surface of the earth and above the ellipsoid = $ d $ \\
\begin{align}
\frac{\partial \gamma}{\partial h} &= -2 \frac{\gamma}{a}(1+f+m-2f(sin^{2}(\phi)))
\end{align}
$m = \omega^2 \dfrac{a^{2}b}{GM}=0.00344978650684$\\
Angular rotation velocity of earth = $\omega = 7.292115 \, 10^{-11} rad./s $  \\
Flattening = $f $ 
\begin{align}
\frac{\partial^2 \gamma}{\partial h^2} &= 6 \frac{\gamma}{a^{2}}\\
\Delta g_{B} &= \Delta g_{F} -\delta g_{B}
\end{align}
 Bouguer gravity anomaly = $\Delta g_{B}$ \\
\begin{align}
\delta g_{B}&=2 \pi G \rho h=1.1195 \, 10^{-6}h
\end{align}
Gravitational attraction of Bouguer plates = $\delta g_{B} $ \\
\begin{align}
\rho &= 2.67 t/m^{3} \\
\delta g_{A} &=.87 \,10^{-5}e^{-.116[(h/1000)^{1.047}]}
\end{align}
Atmospheric correction = $ \delta g_{A} $\\
The height of the Earth surface above the ellipsoid is computed here thanks to the $ GTOPO30$ survey which provides all over the world a resolution of about half a nautical mile.(Credit: U.S. Geological Survey Department of the Interior/USGS)
(see figure ~\ref{fig:height}).

\section{Putting alltogether the corrections (ellipticity, geoid,Bouguer and Free-Air) in the calculation of the relativistic potential for the CERN to Gran-Sasso neutrino delay time correction}
The ellipsoid correction to the General Relativistic delay correction is applied in the following way:

In order to take into account the ellipticity of the earth in the calculation of the relativistic gravitational potential $\Phi$ along the neutrino travel, we have calculated for each point along the neutrino travel inside the Earth, the value of the ellipsoid radius $r_{ellipsoid}(\phi)$ and then the distance to the center $r$ is normalized to this radius of the ellipsoid at this underground point, afterward the relativistic potential is computed to this renormalized point $r\prime$. Some other method can be used (cf. \cite{Dahlen}).
\begin{align}
r\prime &=\frac{r}{r_{ellipsoid}(\phi)} \\
\Phi(r) \Rightarrow \Phi(r\prime) &= k\Phi(\frac{r}{r_{ellipsoid}(\phi)})
\end{align}
$ r :$ distance from the center of the Earth to the point under study along the neutrino road.\\
$ r_{ellipsoid}(\phi) :$ distance from the center of the Earth to the surface of the ellipsoid \\
$ k =1.0000045084305260:$ squared ratio between the surface of the ellipsoid and the sphere of mean radius. \\
The geoid correction to the gravitational potential is applied in such a way that it is maximal at the surface of the ellipsoid and null at the center of the Earth, this is a first order modification, of course more refined correction can improve this calculation.
\begin{align}
&\ \Phi \prime(r,\lambda, \phi^\prime) = \Phi (r,\lambda, \phi^\prime) +2\frac{\zeta(r,\lambda, \phi^\prime)\gamma(\phi)r}{r_{ellipsoid}(\phi)c^{2}}
\end{align}

For the points of the trajectory of the neutrinos which are above the ellipsoid but inside the mountains, we have integrated the gravity acceleration with the Bouguer and the Free-Air correction in order to add a correction term to the gravitational potential $\Phi$ used in the Schwartzchild metric. For the points located below the ellipsoid we have not applied any correction to this relativistic gravitational potential $\Phi$.
The result of the computation (Monte Carlo method with $10^{9}$ iterations ) with all these corrections to the calculation of the delay effect is $ \delta t=4.1863 \,picosecond$ which is equivalent to a length delay of $ 1.2550 mm$, also equivalent to a reduced apparent value of the speed of light of $ \frac{\delta c}{c}=1.7179 \, 10^{-9} $.
\section{Conclusion}
Based on the Schwartzchild metric, a general relativistic metric adapted to the Earth interior is detailed here with geophysics density modelization (PREM). Non rotation and spherical symmetry is assumed to compute an effective zonal prametric general relativistic "gravi-potential" term. The delay of travel for neutrino crossing the Earth from one point of the surface to another distant point is extensively detailed. A Monte-Carlo computation of the integral of the general relativistic time delay effect on neutrino travel from CERN to Gran-Sasso leads to a $\delta t=4.1888 \, picosecond$ in this simple spherical irrotational model. This delay effect is also computed and fitted for other distances crossing the Earth. If a neutrino crossed the whole Earth diameter, a "one inch delay effect" ($2.59 \, cm$) would be observable of $\delta t=86.65 \,picosecond$ . Some secondary effects are later included as the ellipticity of the Earth using the WGS84 system , the GEOID gravitational potential fluctuations using the EGM08 survey and also the Bouguer and Free-Air anomaly gravity corrections using the GTOPO30 survey. The computation of all these corrections leads to a $ \delta t=4.1863 \,picosecond$ delay effect. This general relativistic effect cannot account for the  $\delta t=(57.8 \pm 7.8 (stat.)+8.3-5.9 (sys.)) ns$ seen by the OPERA experiment in Gran-Sasso.
Even the Icarus experiment $\delta t= (0.3 \pm 4.0 (stat.) \pm 9.0s (syst.) ns$ which does not see any discrepancy between the speed of light and the neutrino speed cannot measure such a tiny delay effect due to the general relativistic impact of the mass desnity distribution inside the Earth.

\section{Acknowledgement}
This work was presented at the Meeting on Superluminal Neutrinos APC Paris,12-1-2012.
Because this article is mixing a wide variety of research fields as particle physics, general relativity, geophysics, geodesy, we have chosen to describe as extensively as possible all the calculations carried in this paper, we apologize for those who should disagree with this pedagogical approach. All these computations are made with models and data available at this moment and which seems to us to be sufficient to reach the magnitude of all these effects and of the corrections. Special thanks to Pr. D. Autiero for all the fruitful discussions about the puzzling measurement of the speed of neutrinos with the OPERA experiment. Also I want to thank my colleagues E. Mazzucato, M.Zito, G. Vasseur, S.Emery, B. Vallage, J.Rich and J.M. Levy for critical discussions about the general relativistic effects on neutrino travel inside the Earth.

\end{document}